\documentclass[journal]{IEEEtran}

\usepackage{color}
\usepackage{pdfpages}
\usepackage{cite}
\usepackage{algorithm}
\usepackage{algpseudocode}
\usepackage{amsmath}
\usepackage{multirow}
\usepackage{multicol}   
\usepackage[justification=centering]{caption}
\usepackage{siunitx}
\DeclareMathOperator*{\argmax}{\arg\!\max} 
\usepackage{graphicx}
\usepackage[bottom]{footmisc}

\usepackage{epstopdf,float,amsfonts}
\usepackage{epsfig}

 \title{Resilient  Control of Networked Microgrids using Vertical Federated Reinforcement Learning: Designs and Real-Time Test-Bed Validations}

\author{{Sayak Mukherjee$^{*}$, Ramij R. Hossain$^{*}$, Sheik M. Mohiuddin$^{*}$, Yuan Liu, Wei Du, Veronica Adetola, Rohit A. Jinsiwale, Qiuhua Huang, Tianzhixi Yin, Ankit Singhal}

\thanks{The research described in this paper is part of the Resilience through Data-Driven, Intelligently Designed Control Initiative (RD2C) at Pacific Northwest National Laboratory (PNNL). It was conducted under the
Laboratory Directed Research and Development Program at PNNL, a multiprogram national laboratory operated by Battelle for the U.S. Department of Energy. S. Mukherjee, R. Hossain, S. Mohiuddin, Y. Liu, W. Du, V. Adetola, R. Jinsiwale, T. Yin are with Pacific Northwest National Laboratory, Richland, WA, 99354, USA. Q. Huang is with Electrical Engineering Dept., Colorado School of Mines, USA. A. Singhal is with the Electrical Engineering Dept., Indian Institute of Technology at Delhi, India. Q. Huang and A. Singhal were with PNNL while contributing to this work, 
\textit{Corresponding author: sayak.mukherjee@pnnl.gov}, 
*Equal contributions.
}
}
\begin{document}
\maketitle
\begin{abstract}
Improving system-level resiliency of networked microgrids is an important aspect with increased population of inverter-based resources (IBRs). This paper (1) presents resilient control design in presence of adversarial cyber-events, and proposes a novel federated reinforcement learning (Fed-RL) approach to tackle (a) model complexities, unknown dynamical behaviors of IBR devices, (b) privacy issues regarding data sharing in multi-party-owned networked grids, and (2) transfers learned controls from simulation to hardware-in-the-loop test-bed, thereby bridging the gap between simulation and real world. With these multi-prong objectives, first, we formulate a reinforcement learning (RL) training setup generating episodic trajectories with adversaries (attack signal) injected at the primary controllers of the grid forming (GFM) inverters where RL agents (or controllers) are being trained to mitigate the injected attacks. For networked microgrids, the horizontal Fed-RL method involving distinct independent environments is not appropriate, leading us to develop vertical variant Federated Soft Actor-Critic (FedSAC) algorithm to grasp the interconnected dynamics of networked microgrid. Next, utilizing OpenAI Gym interface, we built a custom simulation set-up in GridLAB-D/HELICS co-simulation platform, named \textit{Resilient RL Co-simulation (ResRLCoSIM)}, to train the RL agents with IEEE 123-bus benchmark test systems comprising 3 interconnected microgrids. Finally, the learned policies in simulation world are transferred to the real-time hardware-in-the-loop test-bed set-up developed using high-fidelity Hypersim platform. 
Experiments show that the simulator-trained RL controllers produce convincing results with the real-time test-bed set-up, validating the minimization of sim-to-real gap.


\end{abstract}

\section{Introduction}
\label{sec:intro}
\subsection{Motivation and Related works}
\IEEEPARstart{I}{n} achieving the goal of decarbonization and net-zero energy by 2050 as highlighted in \cite{netzeroDOE}, the adoption of networked microgrids emerges as a prominent strategy for establishing self-sustaining power grids capable of efficient integration and management of distributed energy resources (DERs). This DERs are commonly interfaced with power-electronic devices, grid-forming/grid-following (GFM/GFLs) inverters \cite{9729134}, the two basic technologies in present day's utility-based IBRs. Commonly, GFL inverters incorporate a phase locked loop (PLL)-based design to track the grid frequency, and operates on a given phase angle regulating the active and reactive power injections, whereas GFMs possess the capability to function as controllable voltage sources linked to a coupling impedance, thereby enabling direct control over the voltage and frequency of the microgrid \cite{8932418,du2020modeling}, and becoming a critical assets of the next generation power grid.


Recent advancements show promising approaches (results) in designing primary controls of GFM technologies \cite{osti_1721727}, but when deployed in a networked microgrid, the control design does not remain limited to the primary level, rather becomes hierarchical with multiple layers  spanning from primary to higher-level, as explored in \cite{bidram2012hierarchical, guerrero2010hierarchical, singhal2022consensus}. That's why these IBRs are becoming vulnerable facing various concerns related to resilience, particularly in scenarios where the power electronic interfaces could be targeted by adversarial cyber attacks, thereby destabilizing the entire power grid. Pertinent literature exploring potential cyber events impacting microgrids and resilience considerations can be found in references \cite{deng2020distributed, zhou2020cyber, sahoo2020resilient}.

On the other hand, in instances involving multi-party ownership models, distinct segments within a networked microgrid might be under the jurisdiction of various utilities/operators. Such settings often involve limited data exchange and proprietary information sharing during operational phases. Furthermore, due to the growing intricacy of microgrid operations and the presence of modeling uncertainties, gaining precise knowledge about the dynamics becomes a challenging endeavor. All these lead us to pose two pivotal research questions: 
\begin{enumerate}
    \item How can we develop higher-level controllers that exhibit efficacy despite having restricted insights on networked microgrids (model complexities, uncertainties, and lack of exact knowledge), thereby enhancing their resilience?
    \item How can we address the issue of limited data sharing across networks of microgrids while accounting for dynamic electrical couplings?
\end{enumerate}

Data-driven solutions are promising avenue to eliminate the need for the exact model knowledge. In particular, Reinforcement learning (RL) has seen considerable progress over last decades solving complex nonlinear dynamic tasks in a Markov decision process (MDP) framework \cite{sutton2018reinforcement} and and can tackle uncertainties up to a certain level. RL optimizes the sequential decision making process using direct interactions with the underlying environment (the system model). 
Different variants of RL, using value-based or policy gradient-based or a combination of both, can be found in literature \cite{LillicrapHPHETS15}. Additionally, RL problems face challenges optimizing tasks over multiple agents in a coupled dynamic environment with segregated action and state spaces; this led to the researches on multi-agent RL (MARL) such as \cite{marl1_review, marl4_maddpg}. In power systems, RL has been utilized for short-term transient voltage control \cite{others_voltage,9442908}, microgrids energy storage control \cite{others_micro}, wide-area damping control \cite{others_wadc}, distribution grids volt-VAr control \cite{others_voltvar}, load frequency regulation problems \cite{yan2018data}. The applications of MARL in power systems can be found for energy management problem \cite{ahrarinouri2020multiagent}, cooperative frequency control \cite{yan2020multi,daneshfar2010load}, automatic generation control (AGC) \cite{li2022coordinated}, optimal use of shunt resources \cite{kamruzzaman2021deep}. Besides, comprehensive overview of RL works related to power systems can be found in \cite{r3,r2,r1}. 

Moreover, the learning framework of generic RL (single agent) and MARL do not ensure privacy regarding raw data; therefore pose challenges in learning problem related to networked microgrid problem having proprietorship data. To tackle this data privacy issues, recent studies on federated learning (FL) \cite{li2020review,bonawitz2019towards, li2020federated}, which shares model (neural network) parameters and gradients between the zones or entities instead of sharing raw data, is a promising pathway. Inline with this idea, Fed-RL, a combination of federated learning and RL, has become popular in recent studies \cite{qi2021federated, wang2020federated, zhuo2019federated}. Overall, Fed-RL is in early stages of development, and some recent works are found in power systems applications, Among those, utilizing Fed-RL, \cite{liu2022federated} solves decentralized volt-var control problem, \cite{li2023wind} proposes privacy preserving wind power forecasting method, \cite{lee2020federated} deploys an energy management system for smart homes, \cite{qiu2023federated} introduces a peer-to-peer energy and carbon allowance trading, and \cite{li2023federated} studies physics-informed reward based multimicrogrid energy management. 

In this paper, we primarily focus on the control design problem to improve the overall resilience of a networked microgrid by mitigating the impacts of adversarial actions at the reference signals of primary control loops of the grid-forming inverters (GFM). A recent work \cite{10089185} studies the destabilizing attacks on the primary control loops of IBRs, and proposed RL based defense design. But this work does not consider the data privacy issues from the viewpoints of a networked microgrid. Also, as reported in \cite{10089185}, limited works can be found on this area. To ensure data privacy in RL, we propose a design architecture of implementing \textit{vertically} federated reinforcement learning framework with the multi-party owned networked (coupled) microgrid. The proposed design architecture is implemented and validated with ResRLCoSIM, developed as a part of this work, for the benchmark system IEEE-123 bus test feeder (modified) with three coupled microgrids. {\color{black} Please note the current work is an extension of our recently published conference paper \cite{mukherjee2022enhancing}, where we have shown proof-of-the-concept implementation. In the current work, we improved the modeling of the test systems with more realistic scenarios considering conventional generators, and inverters; most importantly,} as a next step, this research investigates the aspect of transferring trained RL policies to real-time hardware-in-the-loop test-bed simulations, thereby bridging the gap between theoretical advancements and practical applications. Real-time test-bed simulations allow RL policies to be tested and validated in a controlled environment before deployment in the real world, and provide a safe platform to identify potential issues, vulnerabilities, or unintended behaviors of RL policies. This mitigates the risks associated with deploying untested policies directly into operational systems, where failures can have significant consequences. This also helps designers to mitigate the sim-to-real gap, which is the disparity between the performance of policies learned in simulated environments and their effectiveness when deployed in the real world. We have developed the test-bed simulation setup for the same IEEE-123 bus test system in the Hypersim environment \cite{hypersim}, replicating the microgrid model implemented with simulation platform ResRLCoSIM. Finally, the learned control policies are implemented using Python API via the user datagram protocol (UDP) communication architecture.

%


\subsection{Main contributions}
We summarize the main contributions of this paper as follows:
\begin{enumerate}
    \item A purely data-driven method is proposed to design adversarial resilient control for networked microgrids by reinforcement learning approach in presence of multiple agents. 
    \item Data privacy and proprietary issues among different microgrid owners are solved by blending the ideas of federated learning with reinforcement learning. We proposed a vertical variant of Fed-RL algorithm, FedSAC.
    \item A novel open-source software module, \textit{Resilient RL Co-simulation} (ResRLCoSIM) platform is created utilizing Grid simulator GridLAB-D \cite{chassin2014gridlab} and HELICS \cite{palmintier2017design} co-simulation platform. ResRLCoSIM is compatible with the OpenAI Gym \cite{brockman2016openai} interface and can be used with any benchmark RL methods.
    \item \textcolor{black}{A real-time Python-Hypersim co-simulation platform is developed to test the performance of the trained RL-agents in Opal-RT based real-time simulators. For this purpose, the IEEE-123 node test system with generators/inverter models are developed in Hypersim software and the equivalent GridLAB-D model based trained RL-agents are transferred into the Python environment to bridge the Sim-to-real gap.}
\end{enumerate}


\section{Problem Formulation: Resilient Control for Networked microgrid} \label{sec2}
\label{sec.ARS}
We consider a $N$ bus networked microgrid comprising $m$ microgrids with each having its own GFM inverters. 
\subsection{GFM dynamics} 
Following \cite{du2021model}, we model the $i^{th}$ GFM inverter as an AC voltage source with internal voltage $E_i$, and phase angle $\delta_i$ mathematically represented by (\ref{main_eq1}) and (\ref{main_eq2}).
\begin{align} \label{main_eq1}
    \dot{\delta}_i = u_i^\delta, \\
    E_i = u_i^V \label{main_eq2}
\end{align}
Note that $u_i^\delta, u_i^V$ are the frequency and voltage control input signals or reference signals to the inverter. In this work, We utilize the droop-based primary control of the GFM inverters as given in (\ref{droop1}) and (\ref{droop2}),
\begin{align}\label{droop1}
    \omega_i^{ref} = \omega_i^{nom} - m_{Pi} (P_i - P_i^{set}),\\
    V_i^{ref} = V_i^{set} - m_{Qi} (Q_i - Q_i^{nom}). \label{droop2}
\end{align}
where, $P_i$, and $Q_i$ represents the active and reactive power of the $i^{th}$ inverter. $\omega_i^{ref}$ serves as the frequency control input $u_i^{\delta}$ with $P-f$ droop parameters $m_{Pi}, P_i^{set}, {\omega}_i^{nom}$. $Q-v$ droop control has 3 parameters $m_{Qi}, V_i^{set}, Q_i^{nom}$. But, voltage control input $u_i^V$ is obtained indirectly by passing $V_i^{ref} - V_i$ through a proportional-integral (PI) controller, as the main objective is to regulate terminal voltage $V_i$ rather than $E_i$ (see \cite{du2020modeling,du2021model,singhal2022consensus} for more details). 
These device level primary controls make GFM inverters to behave more like a synchronous generator by actively participating in frequency regulation, active power sharing and mitigating problems of circulating reactive power in parallel operation. 
Additionally, the underlying networked microgird model has GFL inverters as well; the details are not discussed here, as we are mainly interested in resilient control design for the GFM inverters. 

\subsection{Why do we need resilient control layers?}
In normal mode, the control dynamics of the GFM inverters will follow (\ref{droop1}) and (\ref{droop2}). But, if the set-points in (\ref{droop1}) and (\ref{droop2}) are manipulated by an external entity (say, an attacker), the corrupted signal will perturb the set-point signal as follows, 
\begin{align}\label{att1}
    P_i^{set} = P_{i-base}^{set}+ P_i^{attack},\\
    V_i^{set} = V_{i-base}^{set}+ V_i^{attack}.\label{att2}
\end{align}
Please note that $P_{i-base}^{set}$ and $Q_{i-base}^{set}$ are the respective base values. Now, a supervisory control layer can be designed, on top of existing primary and secondary controls if present, to mitigate the effects of the attack signal. We refer this supervisory control layer as the resilient control which will modify the corrupted set-point signals in (\ref{att1}) and (\ref{att2}) as follows,
\begin{align}\label{res1}
    P_i^{set} = P_{i-base}^{set}+P_i^{attack} + P_i^{res},\\
    V_i^{set} = V_{i-base}^{set}+ V_i^{attack} + V_i^{res},\label{res2}
\end{align}
Here, we follow certain standard assumptions like attacker has limited budget, and assume that (a) injected adversary signal is bounded, and (b) attack signal can only be injected at a discrete interval. Next, we discuss this resilient control architecture from the perspectives of a reinforcement learning (RL) problem.
\subsection{RL-based resilient control}
The resilient control inputs $u^{res} = [P_i^{res}, V_i^{res}]_{i=1,..,M}$ can be determined as the output of a feedback function of the microgrid measurements. We call the measured quantities as the observations ($O$); hence $u^{res} = f_\theta(O)$, where $f(\cdot)$ is a parameterized control function with parameters $\theta$. To this end, considering model uncertainties, unknown attack signals, fast computational aspects,  data-driven methods are more promising than model-based methods. This motivated us to utilize RL as a solution by casting this resilient control problem in a partially observed Markov decision process (POMDP) setting. RL is sequential decision making process, where an agent (controller) can learn optimal control actions based on the observations generated due to repeated interactions with a given environment. The MDP (or POMDP) formulation can be represented by a tuple $(S,A, \mathcal{P},R,\gamma)$, where, $S:=$ state space, $A:=$ action space, $\mathcal{P} : S\times A\rightarrow S:=$ environment transition function, $R:=$ reward space such that reward $r:S\times A\rightarrow R$, $\gamma:=$ discount factor $\in (0,1)$. The action $a\in A$ is obtained by a policy $\pi:S\rightarrow A$. The optimal policy is derived by solving $\pi^* = \argmax_{\pi} \sum_{t} \gamma^t r_t= \argmax_{\pi} \sum_{t} \gamma^t {\bf r}(s_t,a_t,s_{t+1})$, where $s_t, s_{t+1}, a_t = \pi(s_t), {r}(\cdot)$ are states, next states, actions at time $t$, and reward function respectively. More details can be found in \cite{sutton2018reinforcement}.

In our setting, we emulated inverter attack by adding adversaries to a set of GFM set-points, and observe the system behavior. It is observed that the these injected attacks can make the system unstable (see no control cases in Fig. \ref{f3}) without any remedial actions. Please note that the unknown (or stochastic) nature of the attack signals eliminates the feasibility of any rule-based controllers. Instead, RL-based resilient and adaptive controllers can be deployed to mitigate the effects of the attack signals, but these RL controllers need to be trained. Next, we discuss on the RL-based resilient controller training with the underlying networked microgrid setting.


Without loss of generality, we consider a resilient control design problem for attacks on the voltage set-point as in (\ref{att2}). In general, microgrid dynamics contains many differential and algebraic variables, but here we we focus on a partial set of such variables, particularly, bus voltage magnitudes $V_i(t)$. Please note that $V_i(t)$ are the terminal voltage of the inverters and are not the set-points. This voltage magnitudes can be obtained from measurements and are termed as observation variable $O$. With slight abuse of notation, we substitute $S$ of general MDP setting with $O$.
The Action space $A$ of the RL agents should contain the resilient control inputs $P_i^{res}$, and $V_i^{res}$ of (\ref{res1}) and (\ref{res2}) for individual GFM. Here, we consider a continuous action profile with some practical set-point limiters implemented to keep the inputs within tolerable bounds. To keep sync with voltage set-point attack problem, we only utilize $V_i^{res}$ as the agent actions. Finally, the reward $r(t)$ defines the objective of the control problem considering quality of service (QoS). As we considered a voltage set-point attack problem, $r_t$ is defined as follows:
\begin{gather}
r_t =
\begin{cases}
 - cu_{ivld} \;\; \text{if}\;\; t \leq t_a,\\
   - \sum_i Q_i {\lvert \lvert V_i(t) - V_{{i,ss}}\rvert \rvert}_2.
\end{cases}
\end{gather}
where, $t_a$ is the instant of the adversarial action, $V_i(t)$ is the voltage magnitude for bus $i$ in the power grid at time $t$, and $V_{i,ss}$ is the steady-state voltage of bus $i$ before the attack, $u_{ivld}$ is the invalid action penalty if the DRL agent provides action when the network is not attacked. $Q_i$ and $c$ are weights corresponding to voltage deviation and invalid action penalty, respectively.
\begin{figure}[t]
  \centering
    \includegraphics[width=0.48\textwidth]{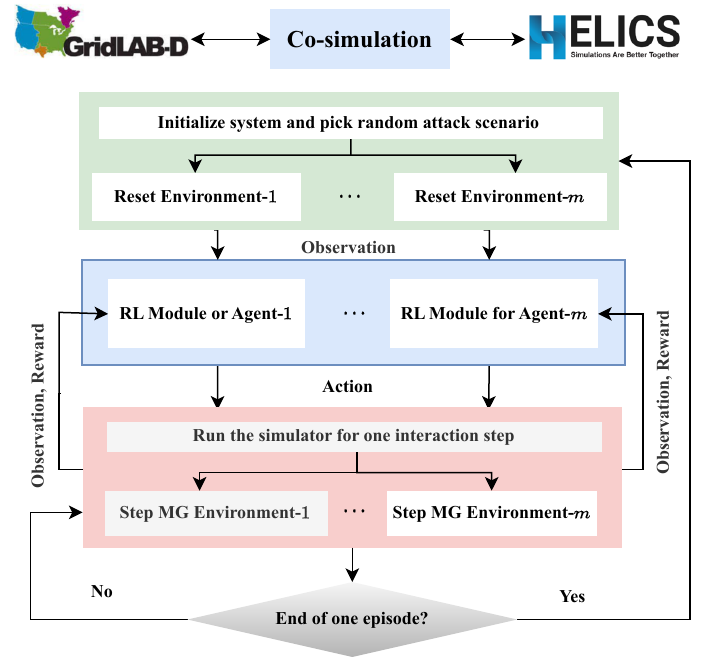}
  \caption{\small{Resilient RL Co-simulation (ResRLCoSIM) Platform for Microgrids}}
  \label{f11}
 \end{figure}
\section{Translation from theory to practice}
This section provides the details of the implementation architecture. First, we discuss the software simulation platform followed by the test-bed integrated hardware-in-the-loop implementation of our proposed RL-based resilient control architecture.
\subsection{Software Simulation Platform: ResRLCoSIM}\label{resrlcosim}

We simulate the microgrid dynamics using distribution grid simulator GridLAB-D. Current research trends in the RL community use OpenAI Gym platform to train benchmark RL algorithms. In line with that, we developed a simulation set-up for microgrids compatible with OpenAI Gym and suitable to train any standard benchmark RL algorithms. To achieve this, the simulation engine needs to be wrapped under a Python API. Plus, we need to implement control tasks (as a part of resilient control schemes) through GridLAB-D. This made us to utilize GridLAB-D's subscription/publication architecture by external customized python codes using the HELICS co-simulation platform. 


Overall, there are two main modules, (a) Co-simulation module enabled by the GridLAB-D/HELICS engine and (b) OpenAI Gym compatible RL algorithm module. Next, we briefly discuss some standard functions associated with OpenAI Gym environment.
\begin{itemize}
    \item \texttt{init()} initializes power flow cases, attack duration, attack instant, and other necessary variables.
    \item \texttt{reset()} makes random selection of necessary configurations including attack signals and start interacting with the GridLAB-D/HELICS module to create a new trajectory roll-out. 
    \item \texttt{step()} establishes the interaction between RL agent (controller) and GridLAB-D/HELICS dynamics. At each step, (a) agent actions are passed to the microgrid, and (b) resulting observations and rewards are returned to the RL module.
\end{itemize}

To conduct the RL training, a pool of adversarial scenarios replicating attacks at primary control loop of GFMs are created. These adversarial attack signal are selected randomly and applied to the system to generate episodic trajectory roll-outs. Finally, the collected episodic trajectory information, including observations, actions, and rewards, are sent to the RL module for training of the RL agent. The detailed framework is shown in Fig. \ref{f11}, which we later utilized for training of our proposed method.
\begin{figure*}[t]
  \centering
    \includegraphics[width=0.85\linewidth]{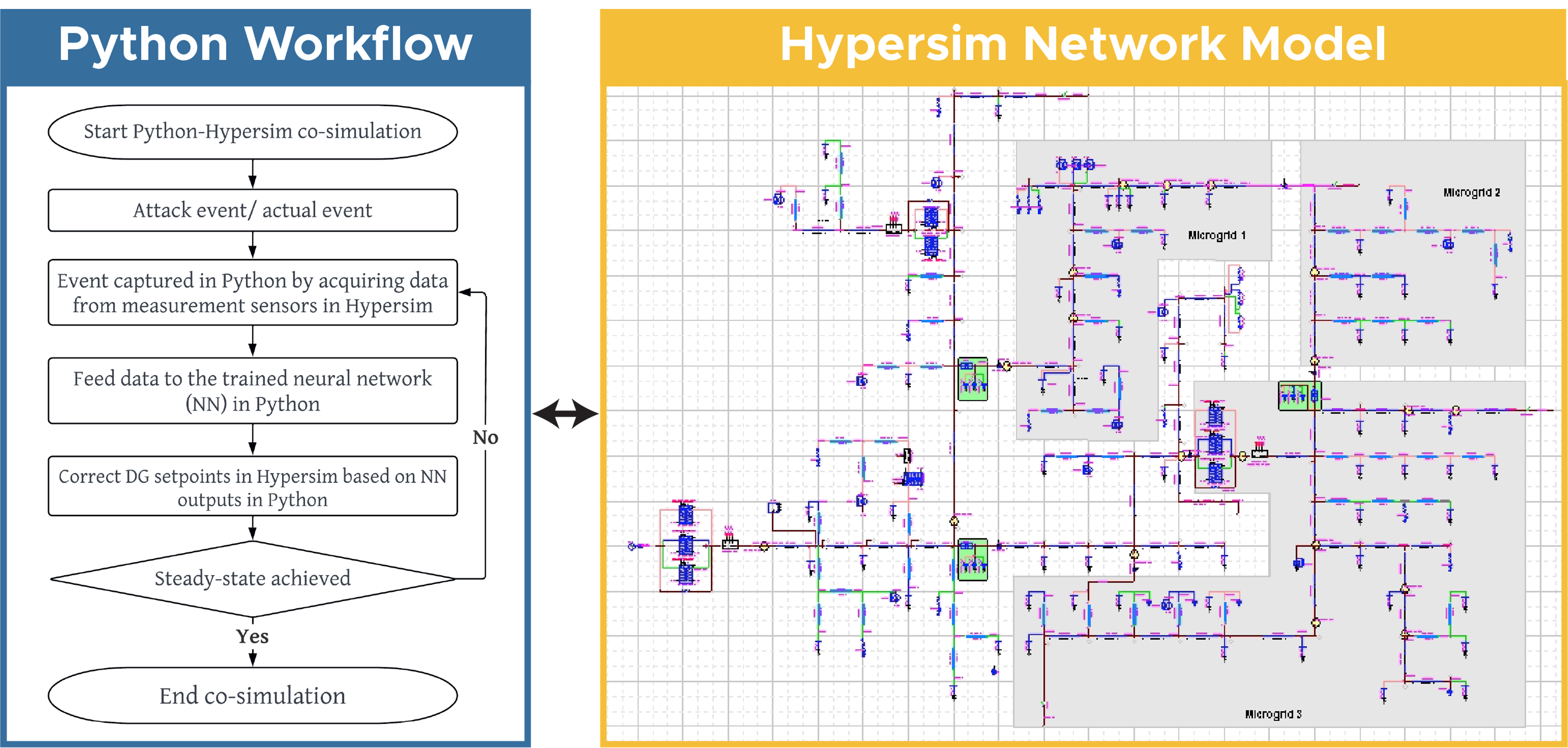}
  \caption{\small{Schematic representation of the Python-Hypersim test-bed co-simulation platform.}}
   \label{fig cosimulation}
 \end{figure*}
\subsection{Sim-to-real Transfer: Hardware-in-the-loop test-bed implementation}
To fill the gap between simulation and hardware-in-the-loop implementation, here we present the architecture to test the performance of the trained resilient RL agents in the real-time simulation platform. For any numerical testing, in this work, we utilize a modified version of IEEE-123 node test systems. Resilient RL agents are trained based on the simulation of this IEEE-123 node test system in ResRLCoSIM platform of Section \ref{resrlcosim}. Therefore, an equivalent IEEE-123 node test system is developed in Hypersim environment for real-time performance evaluation of the RL-agents. The synchronous generators, GFM inverters, and GFL inverters models and controllers in the GridLAB-D and Hypersim simulation platforms are made identical to ensure that the dynamic performances of the generators and inverters in these two simulation platforms matches appropriately. After verification of the controller responses, a co-simulation platform between Hypersim and Python is developed using the UDP communication protocol. Then the trained RL agent models are inserted into the Python API to monitor the responses of the inverters in Hypersim and take control actions in the presence of any adversarial scenarios.A schematic description of this implementation is provided in Fig.~\ref{fig cosimulation}. More details about the real-time simulation platform and verification's are given in Section \ref{Secn: Real-time Validation}. 

\section{Proposed method: Resilient Vertical Fed-RL}

\setlength{\textfloatsep}{0.1cm}
\setlength{\floatsep}{0.1cm}
\begin{algorithm}[t]
\caption{Vertical Fed-RL for networked microgrids}
\begin{algorithmic}[1]
\State \textbf{Initialize} actors (or policies) and critics $\pi_{\theta_k}$ and $Q_{\phi_k}$, respectively, for i.e., $Q_{\phi_k}$, and $\pi_{\theta_k}$ for each microgrid $k$.

\For {$eps = 1,2,\dots,n_f$} 
	    \State \textbf{Sample} an adversarial attack scenario from the  \newline \hspace*{1.3em} adversarial action pool.

    \State \textbf{Generate} episodic trajectory data with ResRLCoSIM.
    
    \State For each of the microgrid $k$, use $o_k$, and $u^{res}_k$ to update \newline \hspace*{1.3em}local critic networks $Q_{\phi_k}$.
    \State \textbf{Send} critic $Q_{\phi_k}$ models to central coordinator.
    
    \State \textbf{Perform} information fusion (or aggregation) at the \newline \hspace*{1.4em} coordinator level by an averaging operation, and return \newline \hspace*{1.3em} parameters of the aggregated global critic network \newline \hspace*{1.4em} model to update each local critic $Q_{\phi_k}$. 

    \State \textbf{Perform} gradient updates of actor parameters of $\pi_{\phi_k}$ \newline \hspace*{1.4em} for each microgrid $k$ using the local observations, \newline \hspace*{1.4em} actions and the updated local critic.
\EndFor
\normalsize
\end{algorithmic} 
\label{alg_fedRL}
\end{algorithm}
\setlength{\textfloatsep}{0.1cm}
\setlength{\floatsep}{0.1cm}

\begin{algorithm}[t]
\caption{Federated Soft Actor Critic (FedSAC)}
\begin{algorithmic}[1]
\State Initialize environments $e_k$, policy $\pi_{\theta_k}$ with parameters $\theta_{k}$, two instances of critic $Q_{\phi_k}$ with parameters $\phi^1_k,\phi^2_k$, and empty replay buffer $\mathcal{D}_{k}$ for all $k = 1,\cdots,m$
\State Set target critic parameters ${\phi_{\text{tar},1}} \leftarrow \phi^1_k, {\phi_{\text{tar},2}} \leftarrow \phi^2_k$ for all $k = 1,\cdots,m$.
\Repeat  
\State Observe $o_{k}$, and select action $u_{k}^{res} \sim \pi_{\theta_{k}}(\cdot|o_k)$ for \newline \hspace*{1.4em}all $k = 1,\cdots,m$.
\State Concatenate actions and form $\cup_{k} u_{k}^{res} = u^{res}$. 
\State Execute and observe next state $o'_{k}$, reward $r_{k}$, and \newline \hspace*{1.4em}done signal $d_{k}$ for all $k = 1,\cdots,m$.
\State Store $(o_k,u_{k}^{res},r_k,{o'}_k,d_k)$ in replay buffer $\mathcal{D}_k$ for \newline \hspace*{1.4em}all $k = 1,\cdots,m$.
\State If ${\cap_{k} d_k} \rightarrow$ TRUE, reset environment state.
\If {Update step is True}
\For {$k = 1,2,\dots,m$} 
 \State Randomly sample a batch of transitions,  \newline \hspace*{4.4em}$B_{k} = \{(o_k,u_{k}^{res},r_k,{o'}_k,d_k)\}$ from $\mathcal{D}_{k}$.
 \State Compute targets for the $Q^k$ functions, (where\newline \hspace*{4.4em} $\tilde{u}_{k}^{res} \sim \pi_k(\cdot|o'_k)$)
   \vspace{-0.05in}
   
  \small
 \begin{equation} \nonumber 
     y_k = r_k + \gamma(1-d_k) \Big ( \min_{i=1,2} Q_{\phi_{k}^{\text{tar},i}}(o'_k,\tilde{u}_{k}^{res}) - \zeta \log \pi^k(\tilde{u}_{k}^{res}|o'_k) \Big )
 \end{equation}
 \normalsize
 \State Update $Q$ functions using: 
   \vspace{-0.05in}
  \small
 \begin{equation}\nonumber 
 \nabla_{\phi_k^i} \frac{1}{|B|} \sum_{(o_k,u_{k}^{res},r_k,{o'}_k,d_k)\in B_k} \Big (Q_{\phi_k^i}(o_k,u_{k}^{res})-y_k\Big)^2,i = 1,2.
 \end{equation}
  \normalsize
  \State Update policy:
  \vspace{-0.05in}
  \small
  \begin{multline} \nonumber 
  \nabla_{\phi_k^i} \frac{1}{|B|} \sum_{o_k\in B_k} \min_{i=1,2} \Big(Q_{\phi_k^i}(o_k,\tilde{u}^{res}_{\theta_k}(o_k)) - \\ \zeta \log \pi^k(\tilde{u}^{res}_{\theta_k}(o_k))|o'_k) \Big )
  \end{multline}
  \normalsize
  \State Update target networks: 
  \begin{equation}\nonumber 
      {\phi_{k}^{\text{tar},i}} = \rho {\phi_{k}^{\text{tar},i}} + (1-\rho) {\phi_{k}^{i}}, \;\;\text{for}\;\; i = 1,2.
  \end{equation}
\EndFor
\If {federated update step}
    \State Compute federated average for critic and target \newline \hspace*{4.4em}$\text{for}\;\;\; i=1,2$.
 \begin{gather*}\nonumber
        {\phi_{\text{fed}}^{i}} = \frac{1}{m}\sum_{k=1}^{m} {\phi_{k}^{i}}\;\;,\;\; {\phi_{\text{fed}}^{\text{tar},i}} = \frac{1}{m}\sum_{k=1}^{m} {\phi_{k}^{\text{tar},i}}
    \end{gather*}
\EndIf
\State Federated update:  ${\phi_{k}^{i}} = {\phi_{\text{fed}}^{i}}$, and 
${\phi_{k}^{\text{tar},i}} = {\phi_{\text{fed}}^{\text{tar},i}}$, \newline \hspace*{3.0em}for $i=1,2$, and for all $k = 1,\cdots, m$.
\EndIf
\Until Convergence
\normalsize
\end{algorithmic} 
\label{alg_fedSAC}
\end{algorithm}
\normalsize
\setlength{\textfloatsep}{0.1cm}
\setlength{\floatsep}{0.1cm}

Federated learning (FL) is a machine learning (ML) branch to train ML models protecting data privacy, security, and proprietary information in presence of multiple entities or parties \cite{li2020review,bonawitz2019towards, li2020federated}. Likewise, to ensure privacy of participating agents, federated reinforcement learning (Fed-RL) is becoming interesting research direction in recent times \cite{qi2021federated}. Fed-RL can be divided into (a) Horizontal, and (b) Vertical versions, where the former one is more conventional but is not appropriate for the underlying microgrid problem as networked microgrids show coupled dynamics. Therefore, in a network of $m$ coupled microgrids, $k^{th}$ microgrid dynamics is not independent of the $j^{th}$ microgrid dynamics. Let us consider, $o_k, u_k^{res}$ represents the observation (bus voltages) and control actions (at GFM set-point) of $k^{th}$ microgrid, respectively. Now, the observation and control actions for the networked microgrid (as a whole) are concatenation of individual microgrid observations and actions, are $O = \cup_k o_k$ and $u_{res} = \cup_k u^{res}_k$, respectively; now due to interconnections the observation $o_k$ depends not only on $u^{res}_k$ but on whole action set $u^{res}$. 
Here, we consider each microgrid has its own RL control agent, where the control policy of $k^{th}$ agent is represented by $\pi_k(\cdot)$, such that $u^{res}_k = \pi_k(o_k)$ for $k=1,\cdots,m$. In this work, we used actor-critic variants of RL algorithm where parameterized neural network (NN) policy $\pi_{\theta_k}$ represents \textit{actor}, while a second parameterized NN $Q_{\phi_k}$ represents the \textit{critic}.

\begin{figure}[t]
  \centering
    \includegraphics[width=0.48\textwidth]{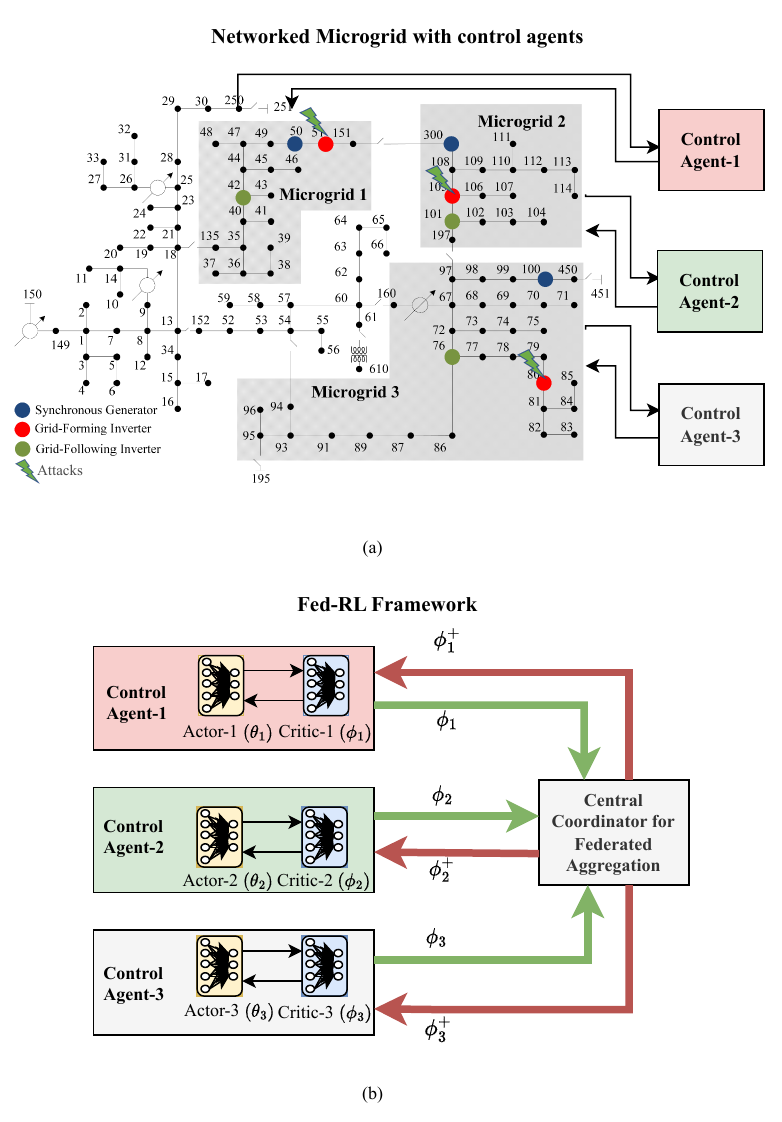}
  \caption{\small{(a) Networked Microgird with individual control agents, (b) Fed-RL framework}}
  \label{f12}
 \end{figure}
 
We infuse the idea of federated learning with the architecture of the actor-critic RL algorithm. As mentioned, the underlying problem considers multi-agent set-up where each microgrid only observes its own measurement data and decides the action of its own agent. But, due to coupled dynamics their observations and actions are inter-dependent, impacting each other through network equations. To deal with this, we utilize the critic network $Q_{\phi_k}$ to achieve the task of federated RL training. First, local microgrid-wise decentralized observation data are utilized to update the local critic networks $Q_{\phi_k}$. After this, the local critic models are sent to a centralized coordinator, for instances operator control center. Please note that in this process microgrids are not transferring any raw data, rather only the critic NN parameters are transferred. Any standard encryption/privacy-preserving techniques can be followed to even prevent any model parameter leakage. But, these data are not that sensitive like raw measurement data. Next, the task of the central coordinator is to aggregate the collected critic models infusing the influence of different microgrid's dynamic behaviors. Like standard federated learning technique, a global (critic) model is initialized and updated using local critic models. To this end, we followed standard federated averaging (FedAvg) technique \cite{mcmahan2017communication}. This aggregated model (or global critic model) is transferred back to individual microgrid, where local critic models are updated with the parameters of the global critic model. Finally, this updated local critic models and local data are utilized to the update the actor or policy NN network $\pi_{\theta_k}$ at microgrid level. This presents a novel multi-agent decentralized privacy preserving RL implementation capturing the coupled microgrid dynamics by the federated averaging of the critic networks. 
The Fed-RL framework follows Algorithm \ref{alg_fedRL}, and Fig. \ref{f12} provides an overview of the comprehensive framework. In this work, we particularly concentrate on the the state-of-the-art soft actor critic (SAC) algorithm \cite{SAC} with entropy regularization, where the algorithm trains the policy maximizing a trade-off between expected return and entropy, which is a measure of randomness in the policy. The standard open source SAC algorithm from Stable-Baselines \cite{raffin2021stable} is extended to incorporate proposed FL framework discussed above. The resulting FedSAC algorithm is presented in Algorithm \ref{alg_fedSAC}.

Next, we discuss some important aspects of Algorithm \ref{alg_fedSAC}.
\begin{enumerate}
    \item In our formulation, bus voltages represent the observation space $o_k$ of individual microgrid, and it is quite natural that steady-state bus voltages are not same. Please note that the steady-state voltages are the solution of power flow equations. Therefore, the values of $o_k$ follow different distribution for different microgrid agents depending on many network factors. Consequently, the state-action space for each microgrid RL agent varies. It is important to note that individual local critics and target critics are trained on local observations, and this distribution mismatch can severely affect the averaging operation of critic and target critic at step 18 in Algorithm \ref{alg_fedSAC}. To solve this issue, we follow microgrid level normalization, where the observation of each individual microgrid are normalized with respect to their steady-state values. 
    \item The standard algorithm of SAC follows a concurrent learning of a policy $\pi_{\theta_k}$ and two Q-functions $Q_{\phi_k^1}, Q_{\phi_k^2}$. In stable baseline \cite{raffin2021stable} implementation, this is conducted following Clipped Double Q-trick, a variant on Double Q-learning that upper-bounds the less biased Q estimate $Q_{\phi_k^1}$ by the biased estimate $Q_{\phi_k^2}$. Usually, a minimum over two Q estimates in step-12 and step-14 of Algorithm \ref{alg_fedSAC} is taken to achieve this. In our experiments, we found that our FedSAC algorithm fails to converge a stable reward value even after promising performance at the initial stage of the training. Further investigation found that at the later part of the training, the weight averaging of the critic and target network (at step 18 of Alg. \ref{alg_fedSAC}) and subsequent Clipped Double Q-trick based minimization operation (step-12 and step-14 of Algorithm \ref{alg_fedSAC}) is detrimental for actor update. But, we also observed the need for the Clipped Double Q-trick at the initial part of training (when actor and critic networks are not still random). To mitigate this issue, we kept the Clipped Double Q-learning for the first half of iterations, after that we select only one critic/target pair either $\{{\phi_{k}^{1}},{\phi_{k}^{\text{tar},1}}\}$ or $\{{\phi_{k}^{2}},{\phi_{k}^{\text{tar},2}}\}$ for federated averaging and actor-critic update. 
\end{enumerate}

\section{Experiments in Simulation platform}
\label{sec.results}
Our propsoed method is implemented with the standard IEEE 123-bus test feeder system \cite{singhal2022consensus}.
The dynamic simulation is performed using our developed ResRLCoSIM platform. In the RL training and testing process, the control agents send action commands to the grid and utilize the observations as feedback. The customized OpenAI Gym interface is utilized to perform RL training. The modified IEEE 123-bus test network consists of three microgrids (MG) with the coupling via tie-lines. The individual microgrids are equipped with 1 GFM inverter, 1 GFL inverter and 1 synchronous generator with rating 600 kW, 350 kW and 600 kVA, respectively with a 
total peak load of 3500 kW. The inverters follow $1\%$ frequency droop and $5\%$ voltage droop values. The GFM inverters are connected at buses (or nodes) 51, 105 and 80 for MGs 1, 2, and 3, respectively while the GFL inverters at buses (or nodes) 42, 101, and 76. Three-phase bus voltages of GFM and GFL inverters are considered as the observations for the underlying RL problem implying $|O| = 6\times 3 = 18$. Now considering multi-agent structure for the Fed-RL problem $|o_{k}| = 3\times 2=6$, for $k = 1,2,3$, as each MG has 2 inverters (1 GFM + 1 GFL).

\begin{figure*}
    \centering
    \begin{minipage}{\linewidth}
    \includegraphics[width = \linewidth]{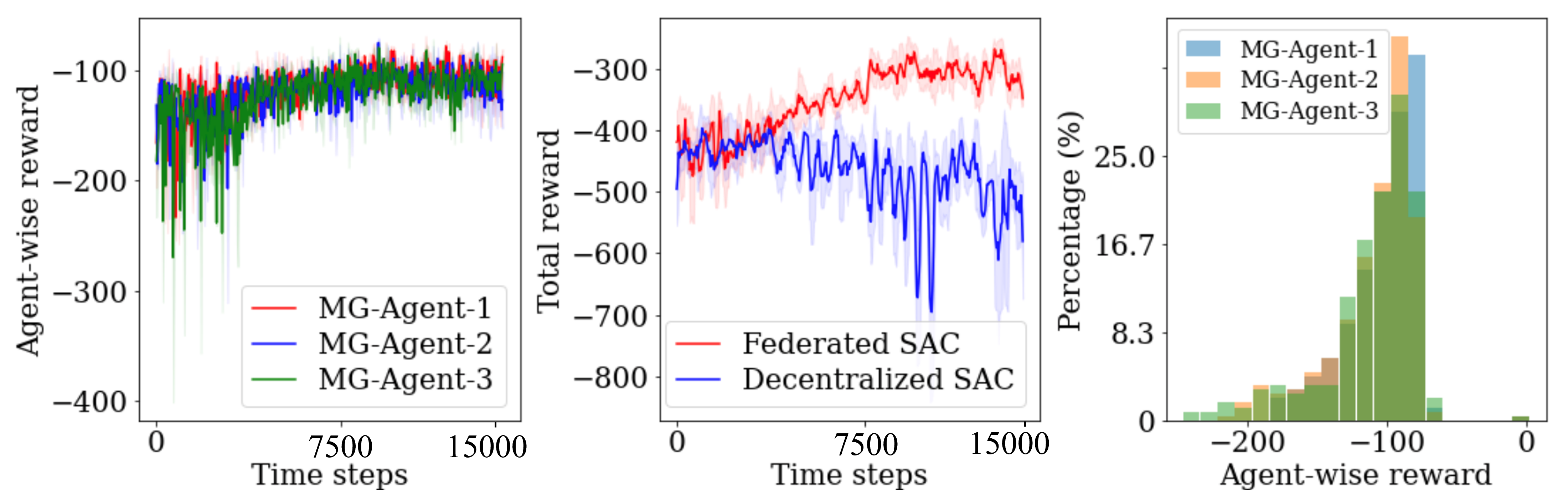}
     \caption{\small{(a) Agent-wise reward plot for Federated SAC training for different seeds, (b) Comparison of Federated SAC and Multi-agent Decentralized SAC, (c) Agent-wise accumulated rewards.}}
        \vspace{0.2 cm}
   \label{f2}
    \end{minipage}
    \begin{minipage}{\linewidth}
        \centering
 \includegraphics[width = \linewidth]{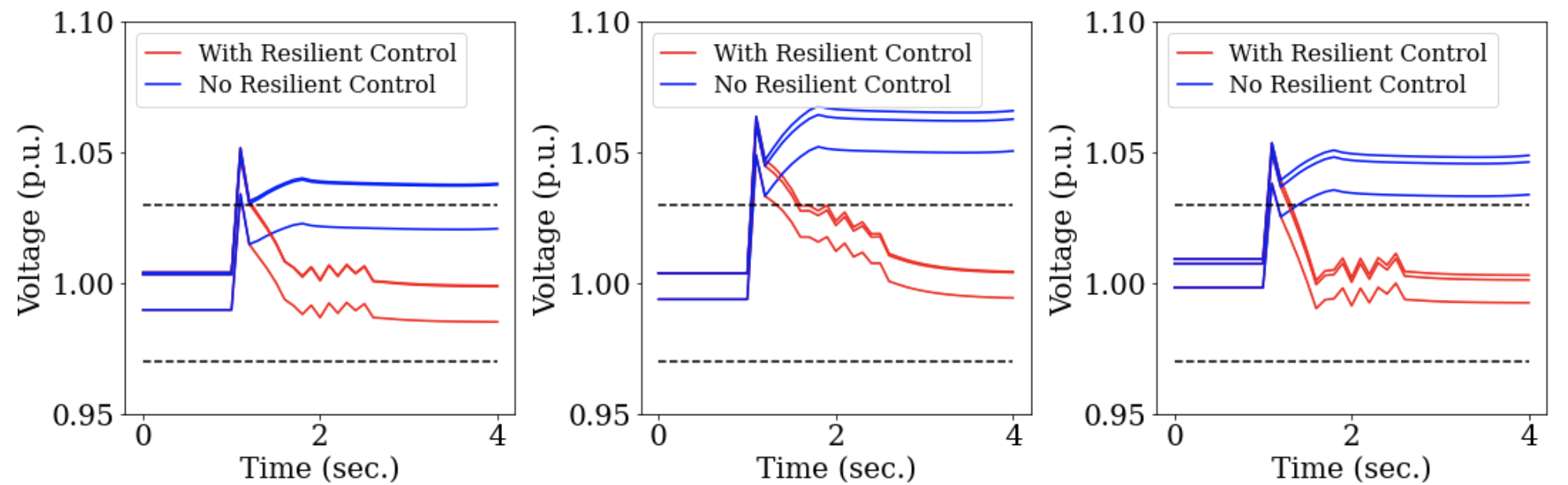} 
    \caption{\small{Testing performance on the grid-forming inverter terminals}}
    \vspace{0.2 cm}
  \label{f3}
    \end{minipage}
     \centering
     \quad
    \quad
    \begin{minipage}{\linewidth}
        \centering
        \includegraphics[width = \linewidth]{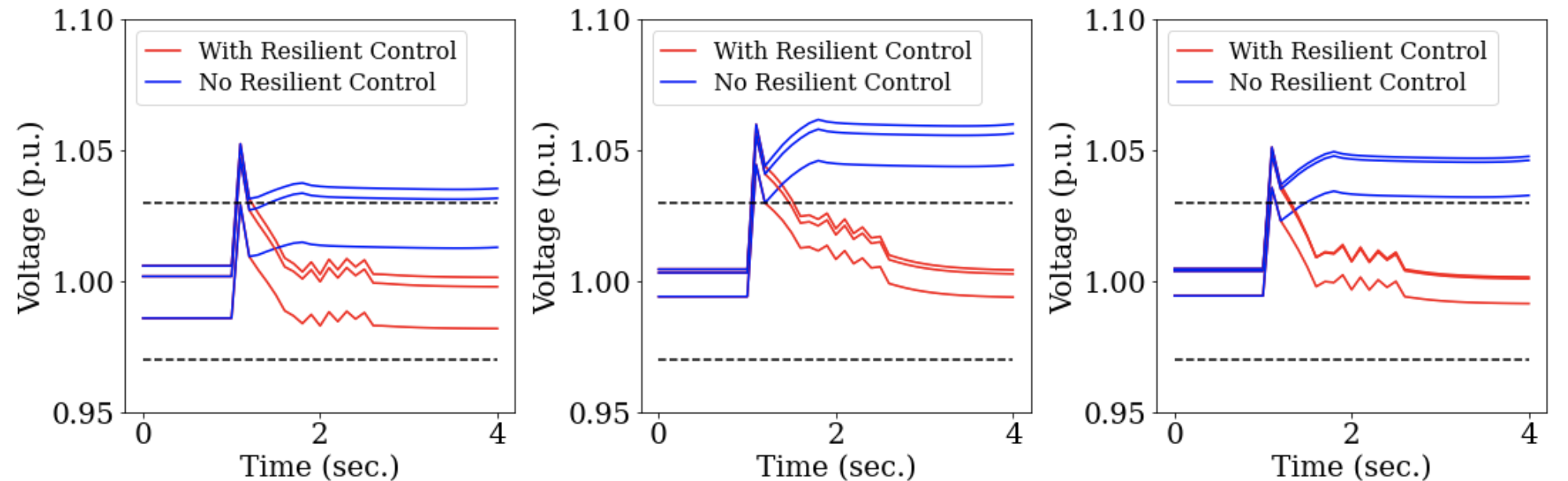} 
        \caption{\small{Testing performance on the grid-following inverter terminals}}
  \label{f4}
    \end{minipage}
    \quad
   \vspace{-0 cm}
\end{figure*}


\begin{figure*}[h]
  \centering
    \includegraphics[width=\linewidth]{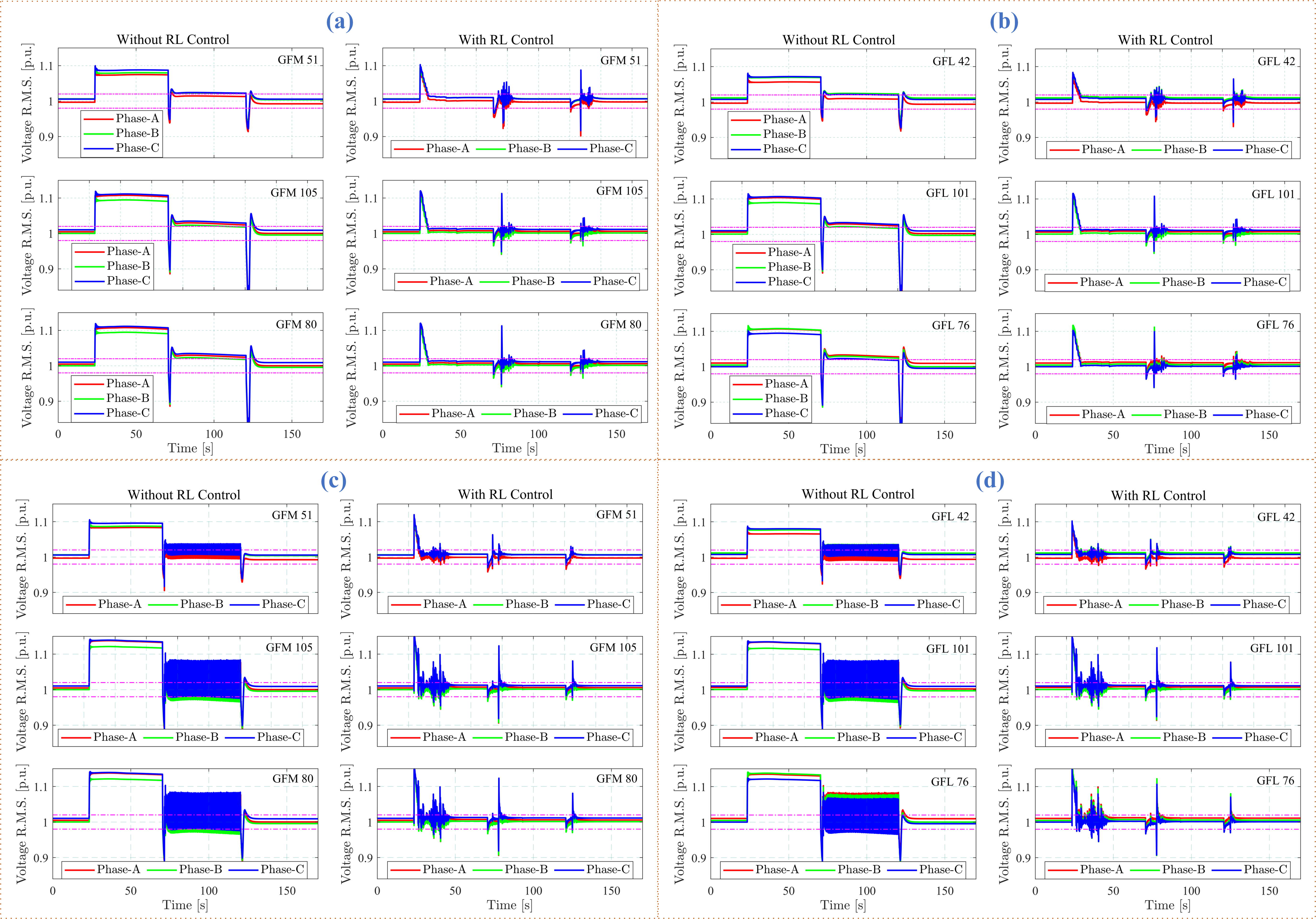}
  \caption{\small{Controller performance during attack on the GFM inverters voltage set-points. (a) GFM inverters response with attack on GFM 51 and GFM 105, (b) GFL inverters response with attack on GFM 51 and GFM 105, (c) GFM inverters response with attack on GFM 51, GFM 105, and GFM 80, and (d) GFL inverters response with attack on GFM 51, GFM 105, and GFM 80.}}
   \label{fig response}
 \end{figure*}

We first need to create a collection of adversarial perturbations. Thereby, we inject attacks at the voltage reference commands for a set of GFM inverters. We perform these events for an episode of $40$ time steps. One of the GFM inverter actuation has been made malicious. For training, we create $7$ different attack scenarios for such GFM actuation points. As described in Algorithm 1 and 2, the FedSAC algorithm has been implemented. The voltage reference points are attacked in the implementation. Voltage set point $V_i^{set}$ of GFM inverters of the respective MG are selected as the actions of the MG agents. Both the actor and critic architectures of each MG-Agent consist of two hidden layers, each containing 64 neurons, and utilize the ReLU activation function. The SAC algorithm is configured with the following training parameters: a learning rate of $0.0003$, a buffer size of $1000000$, a batch size of $256$, $\rho$ set to $0.005$, and $\gamma$ set to $0.99$. The federated learning process commences after 100 time steps and operates at intervals of 10 time steps. In Figure \ref{f2} (a), the training performance of FedSAC is depicted for three distinct microgrid agents, with mean and standard deviations plotted for multiple seed values. Additionally, we conducted experiments comparing the proposed Fed-RL design to a fully decentralized architecture, revealing superior training performance for Fed-RL in Figure \ref{f2} (b). {\color{black} Please, note that to avoid numerical issues in the learning process, we utilized simple action filters based on the voltage observations.} To this end, we conduct tests involving a total of 300 distinct adversarial perturbation scenarios and gather reward data for three different microgrids. We use this data to construct a histogram, as illustrated in Figure \ref{f2} (c). This histogram exhibits a concentration of high reward values, denoting perfect recovery, along with a lower frequency of lower rewards towards the tail, indicative of a high success rate. Figures \ref{f3} through \ref{f4} demonstrate the successful restoration of voltage levels at the selected buses by FedSAC, well within the defined recovery threshold. In contrast, the nominal microgrid model without the resilient controller fails in this regard, thus confirming the efficacy of our design.


 \section{Real-time Hardware-in-the-loop Validation}
 \label{Secn: Real-time Validation}
 \subsection{Python-Hypersim co-simulation platform}
 The synchronous generators, GFM/GFL inverters, and the network components in GridLAB-D simulation platform are solved using the phasor based assumptions which may not capture all the details about dynamic responses due the fast acting controls and high frequency switching components ~\cite{Rohit_Hypersim}. To address this issue, the performance of the proposed resilient RL controller is tested and validated through detailed electromagnetic transient (EMT) simulation in Hypersim-based real-time simulation platform. For this purpose,   
 a co-simulation platform between the Python and Hypersim software's are developed using the UDP communication protocol. 
 Fig. \ref{fig cosimulation}, shows the schematic of the co-simulation platform where the physical systems like IEEE-123 node power system network, synchronous generators,  inverters, loads, and their controllers are modeled inside the Hypersim platform. The Hypersim platform is simulated at $\SI{50}{\micro \second}$ time steps. To perform the real-time simulation at such a lower time-step in Hypersim, the network decoupling approach as presented in ~\cite{Rohit_Hypersim} is used.
 
 In the co-simulation platform, the trained RL controller is imported inside the Python environment. The left-side of Fig. \ref{fig cosimulation} shows the workflow in the Python software which basically takes the measurements from the Hypersim at a predefined time interval and then feds those measurements to the RL controller. Based on the measurements received from the Hypersim, the RL controller generates set points for the inverters which are then fed back into the inverter controllers. In this way the RL controller monitors the status of power system network and take corrective actions in the presence of any adversarial scenarios.

\subsection{Real-time performance evaluation}
This section presents the performance of the inverter controllers through real-time simulations. For this purpose, the voltage set-points of GFM inverters in Microgrid 1-3 are intentionally altered to different levels at (\SI{20}{\second},~\SI{70}{\second}, and \SI{120}{\second})~by adding the attack signals.  
 Fig. \ref{fig response} (a) and Fig. \ref{fig response} (b), respectively demonstrates the performance of the GFM and GFL inverters in Mircogrid 1-3 when attacks are injected to voltage set-points of GFM 51 and GFM 105. From Fig. \ref{fig response} (a) and Fig. \ref{fig response} (b), it can be seen that the RL controller can successfully eliminates the impact of the attacks on GFM and GFL inverter voltage responses and bring back the voltages to the desired limits. In Fig. \ref{fig response} (c) and Fig. \ref{fig response} (d), more extreme scenarios are considered where all the GFM inverters in the test network are corrupted by the attack signal. From the GFM and GFL inverter responses under this scenario, it can be seen that, the inverters without RL controllers loses stability when voltage set points in all the GFM inverters are reduced simultaneously in \SI{70}{\second}-\SI{120}{\second}. However, in the presence of RL controller the inverters can override these extreme scenarios and were able to bound the voltages.
 This simulation results clearly demonstrates the efficacy of the proposed RL controllers in mitigating the presence of attacks in Microgrid network. 

\section{Conclusions}
A novel vertical Fed-RL architecture is proposed to mitigate issues of adversarial attacks in networked microgrids. We added a resilient control layer in conjunction with the primary controls of grid-forming inverters, implemented in a multi-agent fashion and trained using novel FedSAC algorithm to recover grid voltage performance within desired bounds. We developed ResRLCoSIM, an OpenAI Gym compatible GridLAB-D/HELICS co-simulation platform to conduct the training and testing of the RL agents. After successful training and testing, the learned RL policies are transferred from simulation world to real-time hardware-in-the-loop test-bed set-up. Extensive experiments have validated the proposed methods both in simulation and test-bed set up. We will continue the development of novel resilient and secured learning algorithms exploring safe RL aspects, secondary level communication failures, and other variations of adversarial attacks.

\bibliography{ref}
\bibliographystyle{IEEEtran}



\end{document}